\documentclass[superscriptaddress,aps,preprintnumbers,amsmath,showpacs,amssymb,prl,nofootinbib,reprint]{revtex4-1}
\usepackage{bm, color} 
\usepackage{amssymb,amsfonts,slashed,amsthm,amsmath,graphicx, soul,comment}
\usepackage{subfigure}
\usepackage{hyperref}
\usepackage{tikz,xcolor}

\newcommand{\gag}{g_{a\gamma'}}

\begin{document}

\title{Multimodal axion emissions from Abelian-Higgs cosmic strings}

\author{Naoya Kitajima}
\affiliation{Frontier Research Institute for Interdisciplinary Sciences, Tohoku University, \\
6-3 Azaaoba, Aramaki, Aoba-ku, Sendai 980-8578, Japan}
\affiliation{Department of Physics, Tohoku University, \\
6-3 Azaaoba, Aramaki, Aoba-ku, Sendai 980-8578, Japan}

\author{Michiru Uwabo-Niibo}
\affiliation{Cosmology, Gravity, and Astroparticle Physics Group, Center for
Theoretical Physics of the Universe, \\
Institute for Basic Science (IBS), Daejeon  34126, South Korea}

\preprint{TU-1283}

\begin{abstract}
We show that axions can be produced from Abelian-Higgs cosmic strings due to the axion-gauge coupling. The strong magnetic field is confined in the string, and the electric field is induced around the moving string, allowing axion productions from the dynamics of cosmic strings. 
Our numerical analysis on the string collision shows that a sizable number of axions can be produced at the reconnection, and further emissions occur from moving kinks afterward.
Large-scale lattice simulations of the string network further reveal multimodal axion emissions in the sense that axions are produced in both the low-energy and high-energy regimes. The former can contribute to the cold dark matter and the latter can be regarded as dark radiation.  
We found that the axion with GeV or heavier mass can explain the current relic dark matter abundance and simultaneously predicts a sizable amount of dark radiation which can be probed by future observations.

\end{abstract}
\maketitle

{\bf Introduction.--}
The axion is a hypothetical pseudoscalar field originally motivated by the strong {\it CP} problem in the Standard Model (SM) of particle physics \cite{Peccei:1977hh,Peccei:1977ur,Weinberg:1977ma,Wilczek:1977pj}. Beyond this motivation, many extensions of the SM, including certain classes of string theory \cite{Svrcek:2006yi,Arvanitaki:2009fg,Cicoli:2012sz,Petrossian-Byrne:2025mto}, naturally predict the existence of axions. 
The axion can be produced in the early Universe through the misalignment mechanism \cite{Preskill:1982cy,Abbott:1982af,Dine:1982ah} and decay of topological defects such as cosmic (global) strings and domain walls \cite{Davis:1986xc,Lyth:1991bb}. For reviews, see Refs.~\cite{Kawasaki:2013ae,Marsh:2015xka,DiLuzio:2020wdo,Choi:2020rgn,OHare:2024nmr}.

The cosmic string is a linelike topological defect associated with spontaneous breaking of $U(1)$ gauge/global symmetry \cite{Kibble:1976sj,Vilenkin:2000jqa}. 
After the symmetry breaking, the cosmic strings form a network, and it evolves following the scaling regime \cite{Kibble:1984hp,Bennett:1985qt,Bennett:1986zn,Bennett:1989yp,Allen:1990tv}. The network loses its energy by the copious production of loops, and the loop decays with the emission of gravitational waves (GWs) or light particles \cite{Vilenkin:1981bx,Vachaspati:1984gt,Vilenkin:1986ku,Garfinkle:1987yw}.
In particular, the local string loses its energy mainly through GW emissions.\footnote{If the mass of the gauge field is smaller than that of the scalar field, the dark photon emission is allowed, and accordingly, the GW emission is suppressed \cite{Long:2019lwl,Kitajima:2022lre,Kitajima:2023vre}.}
Then, the stochastic GW background from cosmic strings has a characteristic scale-invariant feature in its spectrum, and it can be tested by GW observatories \cite{Cui:2017ufi,Caprini:2018mtu,Auclair:2019wcv}.

Specifically, spontaneous breaking of the Peccei-Quinn (PQ) global $U(1)$ symmetry leads to the formation of axion strings. The network of axion strings evolves with the emission of axions \cite{Davis:1986xc,Vilenkin:1986ku,Garfinkle:1987yw}, and thus the axion string can be a primary source of relic dark matter axions in the postinflationary PQ breaking scenario \cite{Yamaguchi:1998gx,Hagmann:2000ja,Hiramatsu:2010yu,Fleury:2015aca,Gorghetto:2018myk,Kawasaki:2018bzv,Martins:2018dqg,Buschmann:2019icd,Hindmarsh:2019csc,Gorghetto:2020qws,Hindmarsh:2021vih,Buschmann:2021sdq,Saikawa:2024bta,Kim:2024wku,Correia:2024cpk,Kim:2024dtq,Benabou:2024msj}.
The spectrum of axions produced from these strings remains the focus of intensive investigation, as the cosmological role of the emitted particles depends on their momenta: relativistic axions would contribute to dark radiation, while nonrelativistic axions could account for the dark matter observed in the present Universe. Typically, axion strings emit low-energy axions that become nonrelativistic.

Axions can be produced through their coupling to gauge fields, typically arising via anomalies. Indeed, the axion-SM photon coupling enables us to probe the axion by experiments, such as the solar axion search, the light shining through wall experiment, and the cavity search \cite{Sikivie:1983ip,Sikivie:1985yu}, all of which have put the stringent bounds on the couplings. See Ref.~\cite{Irastorza:2018dyq} for a review.

In this paper, we identify and quantify for the first time (to the best of our knowledge) the axion production from a network of local cosmic strings based on the Abelian-Higgs model.
If the axion couples to the (hidden) $U(1)$ gauge field, then the dynamics of the string network can source axion particles. Inside the string core, the $U(1)$ symmetry is locally restored, where the magnetic field ($\bm{B}$) is trapped to stabilize the scalar configuration. Then, the motion of the string generates the electric field ($\bm{E}$) perpendicular to the string, and the oscillation or collision of strings can create regions where $\bm{E}\cdot\bm{B} \neq 0$, producing the axions.
We carry out state-of-the-art large-scale lattice simulations with the number of grid points $N^{3}=8192^{3}$ and show that the emission of axions from the network of Abelian-Higgs strings is {\it multimodal}, in the sense that the axion is produced in both the low-energy and high-energy regimes. Therefore, it provides both the cold dark matter and dark radiation simultaneously, which can explain the present relic dark matter abundance and can be tested by the measurement of the effective number of neutrinos.

{\bf Model.--}
Let us consider the Abelian-Higgs model with an axion coupled to the gauge field \cite{Kitajima:2021bjq,Nakagawa:2022knn,Kitajima:2025nml}. 
The Lagrangian is given by
\begin{align}
\begin{split}
\mathcal{L} &= - (D_\mu \Phi)^* D^\mu \Phi - V(\Phi) - \frac{1}{2} \partial_\mu \chi \partial^\mu \chi - V(\chi) \\
&\qquad - \frac{1}{4} F_{\mu\nu}F^{\mu\nu} - \frac{\gag}{4} \chi F_{\mu\nu}\tilde{F}^{\mu\nu}\,,
\end{split}
\end{align}
where $\Phi$ is the (dark) Higgs field, $\chi$ is the axion field, $F_{\mu\nu} = \partial_\mu A_\nu - \partial_\nu A_{\mu}$ is the field strength tensor for the $U(1)$ gauge fields, $A_\mu$, $\tilde{F}^{\mu\nu} = \epsilon^{\mu\nu\rho\sigma}F_{\rho\sigma}/(2\sqrt{-g})$ is its dual, 
$D_\mu = \partial_\mu-ieA_\mu$ is the gauge covariant derivative with $e$ the gauge coupling constant, and $\gag$ is the axion-gauge coupling which has inverse mass dimension.
Specifically, we consider the following potential for the Higgs field, 
\begin{align}
    V(\Phi) = \frac{\lambda}{4}(|\Phi|^2 - v^2)^2\,,
\end{align}
and we assume that the axion is massless in the subsequent discussion, until we discuss the axion dark matter abundance.

In the flat Friedmann-Lema\^{i}tre-Robertson-Walker spacetime, the equations of motion for the Higgs, the gauge field, and the axion can be derived as follows:
\begin{align}
&\ddot{\Phi}+3H\dot{\Phi} -\frac{1}{a^2}D_i D_i \Phi + \frac{\partial V}{\partial \Phi^*} = 0\,, \label{eq:eom_Phi}\\
&\ddot{A}_i+H\dot{A}_i-\frac{1}{a^2}(\nabla^2 A_i -\partial_i \partial_j A_j) -2e {\rm Im} (\Phi^* D_i \Phi) = 0\,, \label{eq:eom_A}\\
&\ddot{\chi} + 3H\dot{\chi}-\frac{1}{a^2} \nabla^2 \chi + \frac{\partial V}{\partial \chi} = \frac{1}{a^3}\gag \bm{E} \cdot \bm{B}\,,\label{eq:eom_chi}
\end{align}
together with the constraint equation,
\begin{align}
    \partial_i \dot{A}_i -2e a^2 {\rm Im} (\Phi^* \dot\Phi) = 0, \label{eq:constraint}
\end{align}
where $a$ is the scale factor, $H$ is the Hubble parameter, $E_i = F_{0i}$ and $B_i = \epsilon_{ijk} F_{jk}/2$ are the electric and magnetic fields, respectively, the overdot represents the derivative with respect to the cosmic time, $t$, and we impose the temporal gauge $A_0 = 0$.

In the situation of our interest, the axion is emitted but it does not backreact to the gauge field. This assumption is valid if the axion is absent initially, and the axion-gauge coupling is small, i.e.,~$\gag v\leq\mathcal{O}(1)$.
Then, we neglect the axion term in the equation of motion and the constraint equation of the gauge field.
The validity of this no-backreaction approximation is assessed by conservative analytic estimates in the Appendix.

If the axion potential is quadratic (or zero), then the axion-gauge coupling can be absorbed by the redefinition of the axion field ($\chi \to \chi/\gag$). Thus, we set $\gag = v^{-1}$ in our numerical analysis.
In addition, we set $\lambda = 2e^2$ throughout the paper. This corresponds to the so-called Bogomol'ny-Prasad-Sommerfeld (BPS) limit, in which there is no interaction between aligned strings.

{\bf Numerical analysis.--}
We have performed numerical lattice simulations to solve Eqs.~(\ref{eq:eom_Phi})-(\ref{eq:eom_chi}).
Those equations are discretized with the lattice gauge formulation \cite{Moriarty:1988fx}, in which the constraint (\ref{eq:constraint}) is automatically satisfied at each step. 
The spatial derivative is discretized with the second-order precision, and field variables are updated with time using the second-order leap-frog method.

\begin{figure*}[tp]
\centering
\includegraphics [width = 17.5cm, clip]{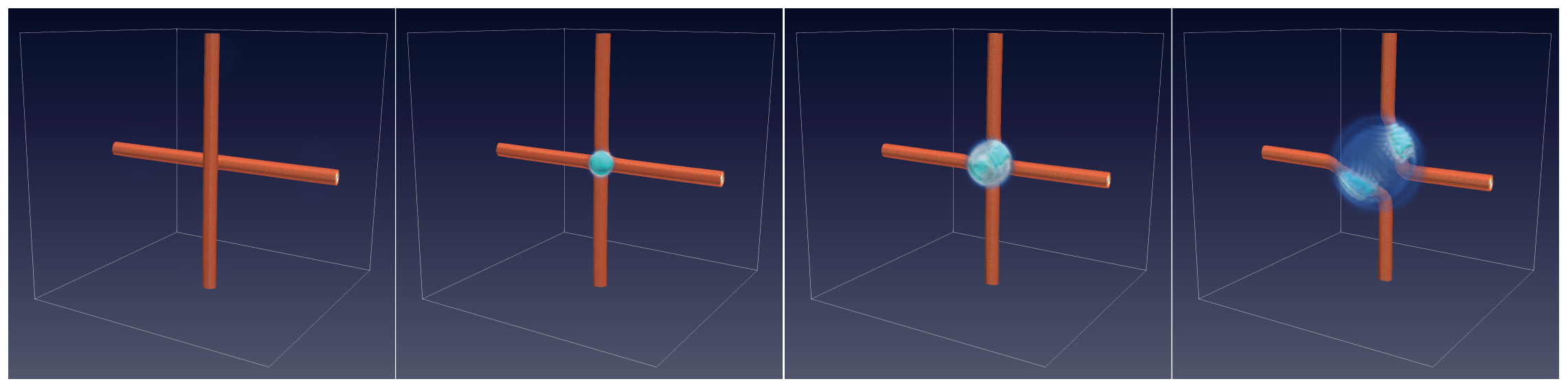}
\caption{
Snapshots of the collision and reconnection of two straight strings (red) and the emitted axions (cyan). $vt = 12.5,\,15.7,\,18.8,\,25.0$ from left to right, corresponding, respectively, to the magenta, cyan, yellow, and thick black lines in Fig.\ref{fig:nk_col}.
}
\label{fig:snapshot}
\end{figure*}

First, we focus on colliding two strings with a right angle in the Minkowski background, as illustrated in Fig.~\ref{fig:snapshot}.
As the initial condition, we prepare two strings, given by the Abrikosov-Nielsen-Olesen vortex solution \cite{Vilenkin:2000jqa}, separated by $d = 16/v$. Then, we perform Lorentz boost so that two strings approach each other with the velocity $v_{\rm str} = 0.5$ \cite{Shellard:1987bv} (see also \cite{Matsunami:2019fss,Hiramatsu:2023epr}).
In this simulation, the number of grids and the box size are $N^3 = 512^3$ and $L = 64/v$ respectively, and we impose the moving-string boundary condition \cite{Hiramatsu:2023epr}.

Four panels in Fig.~\ref{fig:snapshot} show snapshots of the axion emission (cyan) from the collision and reconnection of two strings (red) at $vt = 12.5,\,15.7,\,18.8,\,25.0$. 
It clearly confirms axion radiation from cosmic strings as expected. Remarkably, the axion emission is not restricted to the collision point at the center, but also arises from kinks formed after the reconnection. 

\begin{figure}[tp]
\centering
\includegraphics [width = 8.5cm, clip]{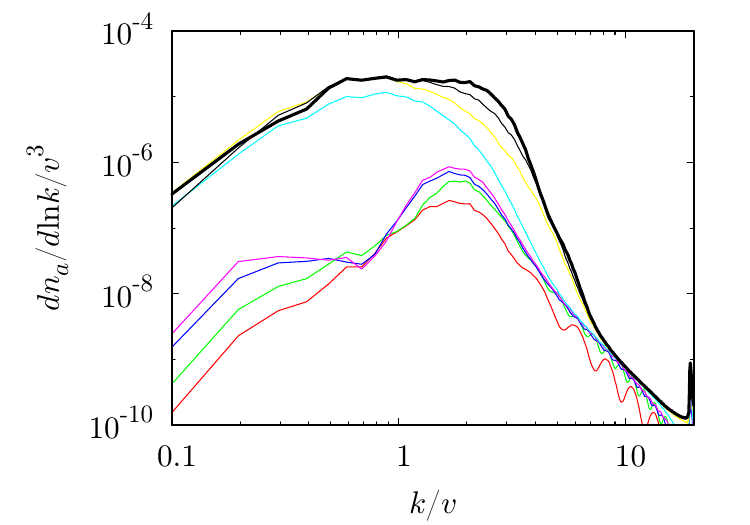}
\caption{
Time evolution of the number density spectrum of the axion emitted from the two string collision. $vt=3.1$\,-\,$25.0$ from bottom to top with the interval $\Delta vt = 3.1$.
}
\label{fig:nk_col}
\end{figure}

Figure \ref{fig:nk_col} shows the time evolution of the spectrum of the number density of emitted axions, $n_a$; $vt=3.1$\,-\,$25.0$ from bottom to top with the interval $\Delta vt = 3.1$ (i.e.,~four panels in Fig.~\ref{fig:snapshot} correspond to the magenta, cyan, yellow, and thick black curves, respectively). 
It exhibits an instantaneous growth of the spectrum from $vt = 15.7$ (magenta) to $vt = 18.8$ (cyan) due to the reconnection of strings. Kinks act as a continuous source of axion radiation, resulting in a further growth of the spectrum.

\begin{figure}[tp]
\centering
\includegraphics [width = 8.5cm, clip]{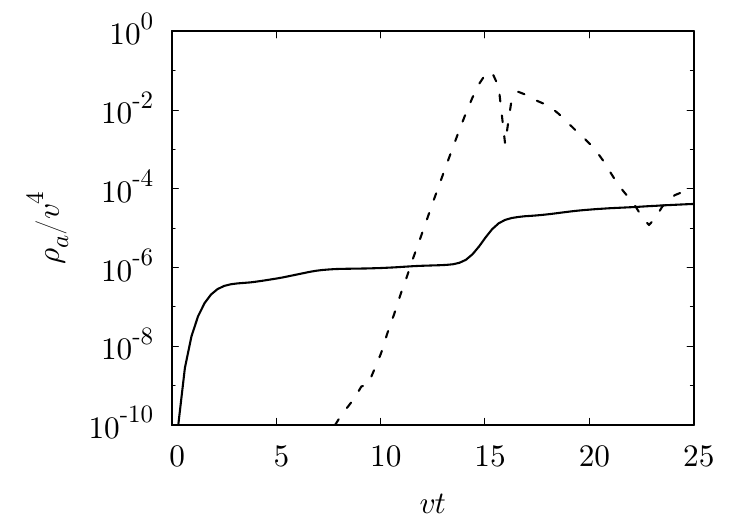}
\caption{
Time evolution of the energy density of the axion. The solid and the dashed lines correspond to the mean value and the value at the center of the simulation box, respectively.
}
\label{fig:rhoa_col}
\end{figure}

The solid curve in Fig.~\ref{fig:rhoa_col} shows the time evolution of the axion energy density averaged over the simulation box. It exhibits a sharp increase around $vt\sim 15$, indicating that the reconnection triggers the burst of the axion.
Note that there is a small amount of axion emissions presented in both Figs.~\ref{fig:nk_col} and \ref{fig:rhoa_col}, even before the string collision. This is because there exists a small but finite magnetic field outside the string core. As the two strings approach, the region with nonzero $\bm{E}\cdot\bm{B}$ is developed even before the collision.

The dashed curve in Fig.~\ref{fig:rhoa_col} shows the evolution of the axion energy at the center of the box (collision point). It remains subdominant compared to 
the energy stored in the string core ($\sim v^4$) even at the collision point, justifying our neglect of backreaction (see Appendix for conservative analytical estimates).
Note that it exhibits two peaks: the first peak appears at $vt\sim 15$, induced by the burst from the reconnection, while the second peak arises when kink-induced axions reach the center.

The change of the intersection angle may suppress the amount of axion produced. The change of the initial string velocity may also alter the production efficiency. 
The systematic study of these dependencies is left for future work.

Next, let us focus on the evolution of the string network in the radiation-dominated Universe.
Here, the number of grid points and the initial box size are $N^3 = 8192^3$ and $L = 128/v$, respectively. The conformal time, $\tau$, is adopted as a time variable.
The initial conformal time is $\tau_i = 1/v$, and the scale factor is normalized as $a(\tau_i) = 1$. We terminate the simulation once the lattice spacing exceeds the string width, at which point the string configuration can no longer be reliably resolved. 
We set the initial value of the Higgs field on the vacuum manifold, $|\Phi| = v$, with a randomly chosen phase at each lattice site.
Other field contents are set as zero.
The simulation starts with the preevolution (relaxation) phase for $\Delta \tau = 2/v$, in which we turn off the axion-gauge coupling because the large gradient energy by the artificial initial configuration may contaminate the signal.
Then, we start the main evolution phase by turning on the coupling, allowing the axion emission from the string network.
We have confirmed that the result is not sensitive to the duration of this preevolution phase.

\begin{figure}[tp]
\centering
\includegraphics [width = 8.5cm, clip]{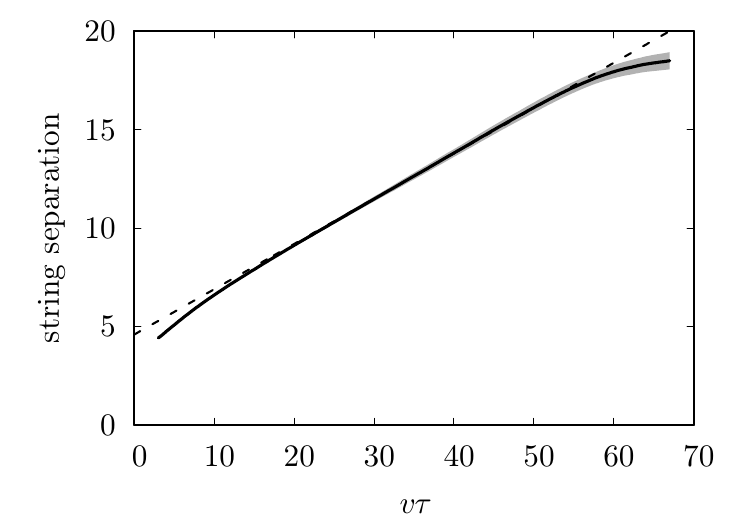}
\caption{
Time evolution of the mean string separation.
}
\label{fig:scaling}
\end{figure}

Figure \ref{fig:scaling} shows the time evolution of the mean string separation defined by 
\begin{align}
    \xi = \sqrt{\frac{L^3}{\ell_{\rm str}}},
\end{align}
where $\ell_{\rm str}$ is the string length estimated by the number of plaquettes with nonzero winding number \cite{Kajantie:1998bg}.
The solid line and shaded region show the mean value and 1-$\sigma$ error from eight independent simulations with different initial conditions. It grows approximately linearly for $10 \lesssim v\tau \lesssim 60$, indicating that the string network has entered the scaling regime during this interval. The deviation from the scaling after $v\tau\gtrsim 60$ is due to the finite-volume effects of the simulation box.

\begin{figure}[tp]
\centering
\includegraphics [width = 8.5cm, clip]{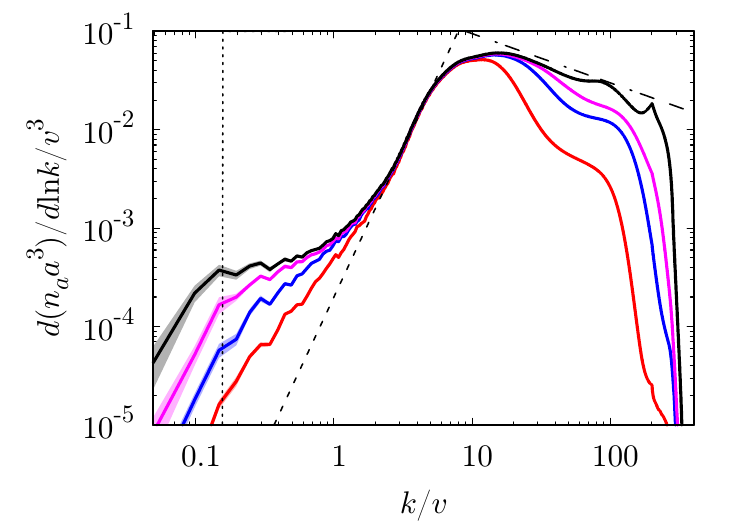}
\caption{
Spectrum of the comoving number density of emitted axions in terms of the comoving wave number, $k$.
Red, blue, magenta, black curves correspond, respectively, to the spectrum at $v\tau = 15,25,35,64$.
The solid line and shaded region show the mean value and 1-$\sigma$ error
from eight independent simulations with different initial conditions.
The dashed and dash-dotted lines show the power-law fitting proportional $k^{3}$ and $k^{-0.5}$, respectively.
The vertical dotted line corresponds to $k/a = 10H$ at $v\tau = 64$.}
\label{fig:spectrum_n}
\end{figure}

Figure \ref{fig:spectrum_n} shows the time evolution of the spectrum of comoving number density of the axion.  
It is observed that the spectrum broadens as time progresses. This behavior likely reflects the continuous production of axions from the string network, resulting in a growth of the number density that surpasses the effect of cosmic dilution.  

In particular, the spectrum exhibits enhancements in both the ultraviolet (UV) ($k \gtrsim 10v)$ and infrared (IR) regions ($k \lesssim v)$, implying the multimodal emissions of high-energy and low-energy axions.
The UV part of the spectrum is primarily induced by string collisions.\footnote{
Production of the massive gauge field is suppressed although it is kinematically allowed \cite{Kitajima:2022lre}.
}
Axions are emitted at a comoving momentum corresponding to the string width, through collisions and kinks formed after reconnections. Since the corresponding $k_{\rm UV} \sim av$ increases with time, the spectrum is gradually extended to higher-$k$ modes.
Peaks around $k \gtrsim 100v$ are due to the numerical artifact.
The spectrum has a positive slope proportional to $k^3$ for $1 \lesssim k/v \lesssim 10$ due to the causality.
The IR region ($k \lesssim v$) can be explained by
the particle production from decaying loops as in the case with the (near-)global string \cite{Gorghetto:2018myk,Long:2019lwl,Kitajima:2022lre} and the semilocal string \cite{Kanda:2025hgi}.

{\bf Dark matter and dark radiation.--}
The broad spectrum of the axion emitted from the string network suggests a potential contribution to both dark matter (DM) and dark radiation (DR). 
Here we provide a semianalytical estimate of the contribution to DM and DR.
For this purpose, we decompose the spectrum into two parts, corresponding to the IR and UV regimes as mentioned above.

Let us first consider the DM axion from the IR regime. The network of long strings loses its energy by chopping off subhorizon scale loops, and the loop decays through the emission of particles and GWs. We assume that the axions are emitted by such loops. Then, the energy density of the emitted axion can be expressed by some fraction of that of the string network, $\rho_a = f  \rho_{\rm str} \propto H^2$, with $f$ a numerical factor smaller than unity.
On the other hand, a typical energy of the emitted axion is $\bar{E}_a \sim O(10) H$, corresponding to the IR peak of the spectrum in Fig. \ref{fig:spectrum_n}.
Then the number density of the emitted axion evolves as $n_a \sim \rho_a/\bar{E}_a \propto H$ (see also \cite{Gorghetto:2018myk,Long:2019lwl,Kitajima:2022lre,Kanda:2025hgi}). 
Denoting $n_a = \epsilon v^2 H$, with $\epsilon$ a numerical coefficient, 
$\epsilon$ reflects the emission efficiency of the axion from loops, incorporating the geometrical suppression of the source term $\bm{E}\cdot\bm{B}$. 
From the numerical result, namely, $n_a a^3 \simeq 4 \times 10^{-4} v^3$ at $v \tau = a = 64$ (see Fig. \ref{fig:spectrum_n}), we obtain $\epsilon \simeq 6 \times 10^{-6} (\gag v)^2$, where the $\gag$-dependence is revived. Note that $\epsilon \sim O(1)$ in \cite{Kanda:2025hgi}.

Assuming that the axion acquires the mass $m_a$,
the production of IR axions stops when the Hubble parameter becomes comparable to the axion mass \cite{Long:2019lwl,Kitajima:2022lre}, and the abundance of the IR axion is fixed at that time. 
Then, the DM axion abundance is estimated as follows,
\begin{align}
\Omega_a^{\rm (DM)} h^2 &\simeq 0.019 (\gag v)^2 \bigg(\frac{g_{*{\rm SM}}}{g_{*s}(T_1)} \bigg) \bigg(\frac{g_{*}(T_1)}{g_{*{\rm SM}}} \bigg)^{\frac{3}{4}} \nonumber\\
&\qquad \times \bigg(\frac{m_a}{1\,{\rm eV}}\bigg)^{\frac{1}{2}}\bigg(\frac{v}{10^{14}\,{\rm GeV}}\bigg)^2\,,
\end{align}
where $T_1$ is the cosmic temperature at $H = m_a$, $g_*$ ($g_{*s}$) denotes the effective relativistic degrees of freedom for the background energy (entropy) density and $g_{*{\rm SM}} = 106.75$ as a reference value.
 
\begin{figure}
    \centering
    \includegraphics[width=8.5cm]{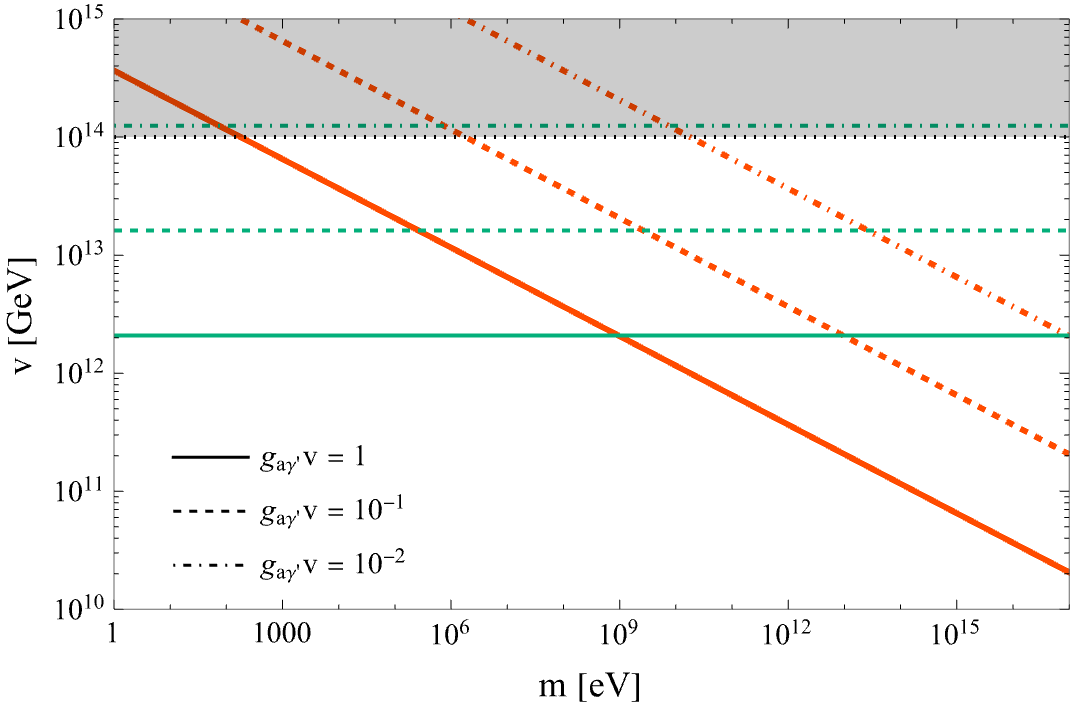}
    \caption{Contour of $\Omega^{\rm (DM)}_a = \Omega_{\rm DM}$ are shown in red for $\gag v=1$ (solid), $\gag v=10^{-1}$ (dashed), and $\gag v=10^{-2}$ (dot dashed), respectively. The green horizontal lines show the corresponding DR constraints $\Delta N_{\rm eff}\leq 1$ at $z=1100$ with $A = 0.06$, $\alpha = -0.5$, and $k_*= 10v$. The gray-shaded region denotes the constraint from the GW observation.
    }
    \label{fig:DM2}
\end{figure}

Figure~\ref{fig:DM2} presents the contours of $\Omega^{\rm (DM)}_ah^2 = \Omega_{\rm DM}h^2$ $\simeq 0.12$ \cite{Tristram:2023haj}, above which the axion DM is overproduced.
The gray-shaded region and the green horizontal lines represent the constraints from the GW observation, $v\leq 10^{14}\,{\rm GeV}$ \cite{Figueroa:2023zhu}, and the DR abundance (see below), respectively.
This result shows that low-energy axions with mass $\gtrsim 1$ GeV produced from the Abelian-Higgs string network can explain the relic DM abundance. This is a novel and viable mechanism for the axion DM production.
\footnote{
In the presence of a kinetic mixing with the SM sector, the axion DM may decay into SM photons. 
$\gamma$-ray observations place constraints on the axion-photon coupling  $g_{a\gamma}\lesssim  10^{-23}\,{\rm GeV^{-1}}$ for masses around $1\,{\rm GeV}$ (corresponding to the lifetime bound $\tau \sim 10^{26}\,{\rm s}$ of the axion DM) \cite{Sun:2023acy}. 
For nonrelativistic axions, the effective coupling induced by the kinetic mixing is typically
$g_{a\gamma}\sim\varepsilon g_{a\gamma'} (m_a^{2}/m_{A'}^{2})$ where $\varepsilon$ is a mixing parameter and $m_{A'}\,(=ev)$ is the gauge field mass. It is extremely small in the case of our interest (e.g. $m_a \sim 1$ GeV and $m_{A'} \sim 10^{13}$ GeV).
}

\begin{figure}[tp]
\centering
\includegraphics [width = 8.5cm, clip]{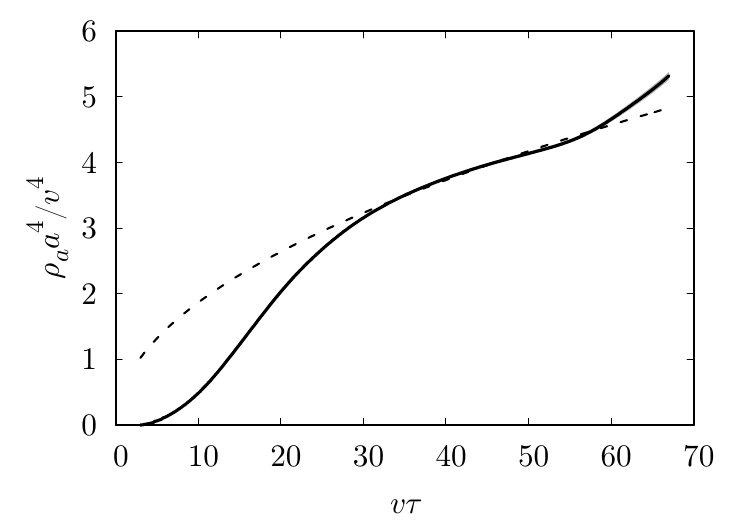}
\caption{
Time evolution of the axion energy density multiplied by $a^4$.
The solid line is the numerical result and the dashed line shows Eq.~(\ref{eq:rho_DR}) with $A=0.06$, $k_*/v=20$, $\alpha = -0.5$ and $\gag v = 1$ multiplied by the numerical fudge factor $1.1$.
}
\label{fig:axion_energy}
\end{figure}

Next, let us consider the DR axion from the UV regime.
The time evolution of the spectrum in Fig. \ref{fig:spectrum_n} indicates a clear tendency to extend to the high-$k$ direction. Then, we postulate that the spectrum takes the following form:
\begin{align}
\frac{d(n_a^{\rm (DR)} a^3)}{d\ln k} = A (\gag v)^2 v^3 \left( \frac{k}{k_*} \right)^\alpha  ~~\text{for}~~ k_* < k < k_{\rm UV},\label{eq:extrapolation}
\end{align}
where $k_*$ denotes the peak wave number, $A$ and $\alpha$ are numerical constants, and $k_{\rm UV}$ is the ultraviolet cutoff scale corresponding to the string width, that is, $k_{\rm UV}/a \sim v$.
The energy density of the relativistic axion can be calculated by integrating the spectrum over $k_* < k < k_{\rm UV}$, which leads to
\begin{align}
\rho_a^{\rm (DR)} &= \frac{1}{a^3}\int_{k_*}^{k_{\rm UV}} d\ln k \,\frac{k}{a} \frac{d(n_a^{\rm (DR)}a^3)}{d\ln k} \nonumber \\[1mm]
&\approx \frac{A(\gag v)^2 a^{\alpha-3}v^4}{(k_*/v)^\alpha (1+\alpha)},
    \label{eq:rho_DR}
\end{align}
where we assume $\alpha > -1$ and $k_{\rm UV} \gg k_*$, and thus ignored the contribution from the lower limit of the integral.
Therefore, the contribution to the effective number of extra neutrinos, $\Delta N_{\rm eff} = c_{\rm eff} \rho_a^{\rm (DR)}/\rho_\gamma$, is
\begin{align}
    \Delta N_{\rm eff} 
    &\simeq 2.9 \times 10^{13+24.9\alpha} \frac{c_{\rm eff} A (\gag v)^2}{(k_*/v)^\alpha (1+\alpha)} \nonumber \\
    &\qquad \times \left(\frac{\rm 1\,eV}{T} \right)^{1+\alpha} \left(\frac{v}{10^{13}\,{\rm GeV}} \right)^{\frac{5+\alpha}{2}}\,,
\end{align}
where $\rho_{\gamma}=\pi^2 T^{4}/15$ is the photon energy density, and $c_{\rm eff}^{-1} = (7/8)\times (4/11)^{4/3}$.
Our numerical result reads $k_* \simeq 10v$, $A \simeq 0.06$, and $\alpha \simeq -0.5$.

Figure \ref{fig:axion_energy} shows the time evolution of the axion energy density in the simulation (solid curve), compared with the semi-analytic formula Eq.~\eqref{eq:rho_DR} (dashed curve). The semianalytic formula reproduces the numerical results well for $v\tau\gtrsim 35$ up to an overall fudge factor (10\% correction), justifying our extrapolation in Eq.~\eqref{eq:extrapolation}. The deviations observed for $v\tau\gtrsim 60$ and $v\tau \lesssim 35$ are attributed, respectively, to the finite-volume effects (see Fig.\ref{fig:scaling})
and the ignorance of the lower limit of the integral in Eq.~(\ref{eq:rho_DR}).

\begin{figure}
    \centering
    \includegraphics[width = 8.5cm, clip]{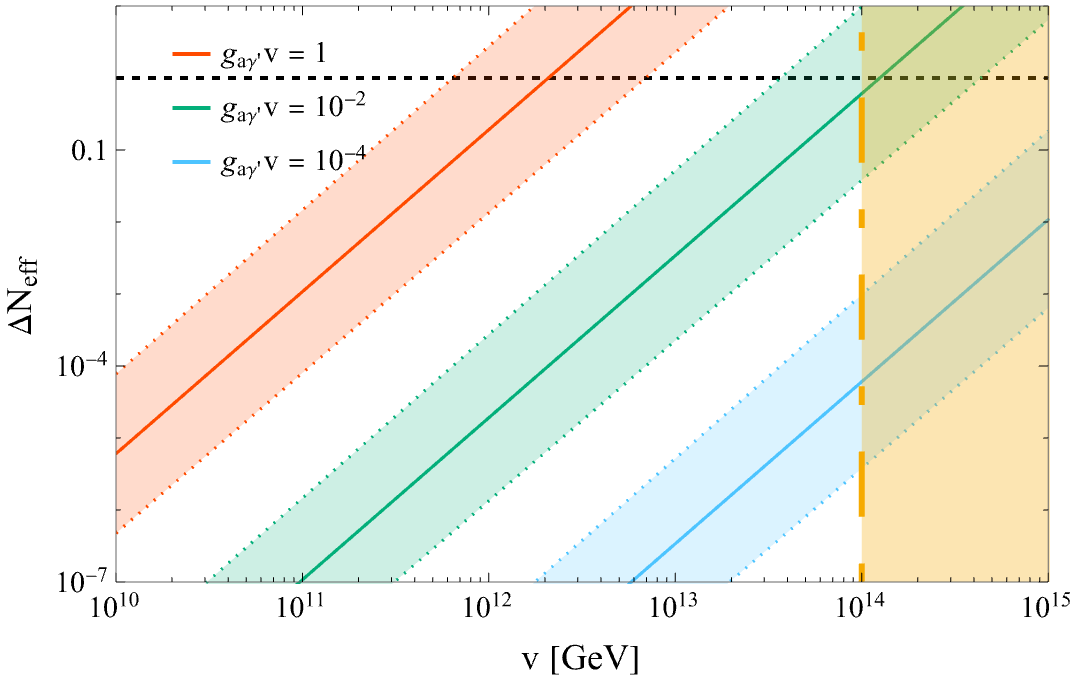}
    \caption{Dependence of $\Delta N_{\rm eff}$ on the symmetry-breaking scale $v$ at the cosmic microwave background epoch ($z=1100$). The cases $\gag v=1, 10^{-2}, 10^{-4}$ are shown in red, green, and blue, respectively. The black dashed line indicates $\Delta N_{\rm eff}=1$ for reference. The yellow shaded region, bounded by the dot-dashed line, is excluded by GW observations. 
    $A = 0.06$, $\alpha = -0.5$, and $k_{*} = 10v$ are adopted
    with shaded bands illustrating $10\,\%$ uncertainty in $\alpha$.}
    \label{fig:DR}
\end{figure}

Figure \ref{fig:DR} shows $\Delta N_{\rm eff}$ at the redshift $z=1100$ for different values of the coupling $(\gag v)$. with $10\,\%$ uncertainty in the tilt-parameter $\alpha$. 
This figure illustrates that axions produced by the Abelian-Higgs string network could potentially be probed through the DR search by the cosmic microwave background or large scale structure observations \cite{Planck:2018vyg, RoyChoudhury:2024wri, Rossi:2014nea}.

Note that produced axions can induce an additional friction on the string network dynamics. However, we neglect such a backreaction in our numerical setup, and thus, it should be checked whether the axion-induced friction is subdominant to the Hubble friction due to the cosmic expansion.
We provide a conservative estimate of the friction force from both IR and UV axion components in the Appendix, 
and show that the backreaction is not efficient in our numerical setup.

{\bf Discussion.--}
We have 
shown a novel mechanism of axion production from Abelian-Higgs cosmic strings, arising from the coupling with the $U(1)$ gauge field.  
Our simulations demonstrate that axions can be efficiently produced from cosmic string collisions, and we further investigated the axion emission from the string network.

We found that the spectrum of the axion emitted from the network exhibits enhancements in both IR and UV regions, and such a broad spectrum suggests sizable contributions to both DM and DR.
By extrapolating the numerical result, we estimated those contributions semianalytically and concluded that the Abelian-Higgs string network can produce a sufficient amount of axion DM, which can explain the present DM relic abundance.
At the same time, the scenario predicts a sizable amount of DR, which can be probed through cosmic microwave background observations.
Current cosmological constraints require the GeV or heavier axion mass, which could potentially be probed by $\gamma$-ray telescopes \cite{Watanabe:2025axr}.
More precise and longer simulations with more grid points and larger box sizes are required for further investigation.

Our numerical results imply that axions are produced by string loops. In other words, it provides a new decay channel of the string loop, which potentially suppresses the GW emissivity and thus significantly changes the resultant GW spectrum.
Moreover, our simulation neglected the backreaction of axion production on the string dynamics. Such effects, particularly relevant for $\gag v \gtrsim 1$, can significantly modify the string network evolution. We leave a detailed study on these issues for future work.

\section*{Acknowledgments}
We thank Fuminobu Takahashi for useful comments. We also thank Wan-Il Park for helpful comments on the arXiv version of this paper.
This work used computational resources of Fugaku supercomputer, provided by RIKEN Center for Computational Sciences, through the HPCI System Research Project (Project ID: hp250177). The work of M.U.~is supported by JSPS Grant-in-Aid for Research Fellows Grant No.~24KJ1118 and by IBS under the project code, IBS-R018-D3.

\appendix
\section{Appendix: Validity of the no-backreaction assumption}\label{app:backreaction}

Throughout this work, we neglect the backreaction from the produced axions on the string dynamics through the axion-gauge interaction. In other words, we drop the axion term in the equation of motion and the constraint equation for the gauge field (see also the setup described in the main text). In this appendix, we provide conservative analytic estimates of the backreaction and justify our numerical setup.

{\bf Backreaction in the collision region.--}
A conservative check is to compare the energy density of the axion with that of the string at the string collision point.
Considering that the axion is locally excited around the string core, 
the local energy density of the axion in this region is
\begin{align}
    \rho_a \sim (\nabla\chi)^2\sim \frac{\chi^2}{\delta^2},
\end{align}
where $\delta \sim (\sqrt\lambda v)^{-1}$ is the string core width.
On the other hand, the typical amplitude of the axion sourced by colliding strings is estimated as $\chi \sim g_{a\gamma'} (\bm{E}\cdot\bm{B})\,\delta^2$,
where the magnetic field is $\mathcal{O}(v^2)$ and the electric field is induced as $E \sim v_{\rm str} B$ around the moving string. Note that the scalar and magnetic core sizes are the same in the BPS limit.
Then, one obtains
\begin{align}
    \chi \sim g_{a\gamma'} v_{\rm str} B^2 \delta^2 \sim  g_{a\gamma'}\, v_{\rm str}\, v^2.
\end{align}
Thus, the ratio of the energy density of the axion to that of the string, $\rho_{\rm str} \sim v^4$, at the collision point is
\begin{align}
    \frac{\rho_a}{\rho_{\rm str}}
    \sim 
    \frac{\chi^2 / \delta^2}{v^4}
    \sim (g_{a\gamma'} v)^2 v_{\rm str}^2.
\end{align}
Therefore, a naive criterion for neglecting the backreaction at the collision is
\begin{align}
    (g_{a\gamma'} v)\,v_{\rm str} \ll \mathcal{O}(1)\,.
\end{align}

{\bf Backreaction on network evolution.--}
The accumulated axion background can induce an additional friction to the string network. Then, we need to check whether this friction remains subdominant to the Hubble friction \cite{Vilenkin:2000jqa}, i.e.
\begin{align} \label{eq:Fdrag-muH}
    \frac{F_{\rm drag}}{L} \ll \mu H \,.
\end{align}
The left-hand side denotes the axion-induced drag force per unit length on a string segment moving with velocity $v_{\rm str}$, which can be roughly estimated as
\begin{align}
    \frac{F_{\rm drag}}{L}
    \sim
    n_a\, v_{\rm str}\, k_\perp \,\frac{\sigma_{\rm tr}}{L},
\end{align}
where $k_\perp$ is the typical axion momentum transverse to the string, and $\sigma_{\rm tr}/L$ is the transport cross section per unit length.
Although $\sigma_{\rm tr}$ is model dependent, we adopt a conservative upper bound based on two-dimensional
(line-defect) unitarity, $\sigma_{\rm tr}/L \lesssim k_\perp^{-1}$ \cite{Vilenkin:2000jqa}, 
which leads to
\begin{align}
    \frac{F_{\rm drag}/L}{\mu H}
    \lesssim
    \frac{v_{\rm str} n_a}{\mu H}.
\end{align}
Then, our requirement (\ref{eq:Fdrag-muH}) reads $n_a/(\mu H)\ll 1$. In the following, we separately estimate the IR and UV contributions
to $n_a$.

Following the parametrization used in the main text, we write the number density of the IR axions as $n_a^{\rm(IR)}=\epsilon\,v^2 H$, where our numerical results give $\epsilon\simeq 6\times 10^{-6}(g_{a\gamma'}v)^2$. Then we obtain
\begin{align}
    \frac{n_a^{\rm(IR)}}{\mu H}
    =
    \epsilon\,\frac{v^2}{\mu}
    \sim
    10^{-6}\,(g_{a\gamma'}v)^2\,,
\end{align}
and our requirement is well satisfied for $g_{a\gamma'}v\lesssim \mathcal{O}(1)$.
For the UV component, we use the semianalytical formula \eqref{eq:extrapolation}, which implies $n_a^{\rm(UV)}a^3 \sim 
A(g_{a\gamma'}v)^2 v^3$ with $A \sim 0.06$. 
Using $H\simeq 1/(a\tau)=v/a^2$, we obtain
\begin{align}
    \frac{n_a^{\rm(UV)}}{\mu H}
    \sim
    \frac{A(g_{a\gamma'}v)^2 v^3/a^3}{\mu\,v/a^2}
    \sim
    \frac{A}{a}\,(g_{a\gamma'}v)^2\,,
\end{align}
with $a \gtrsim 25$ in our numerical setup and hence, 
the UV contribution also satisfies $n_a^{\rm(UV)}/(\mu H)\ll 1$ for $g_{a\gamma'}v\lesssim \mathcal{O}(1)$.
Therefore, both locally (collision region) and globally (network evolution), the above conservative analytic estimates imply that axion backreaction is negligible in the parameter region explored in this work. This supports the no-backreaction assumption adopted in our simulations and the semianalytic extrapolations.

\bibliography{ref.bib}

\end{document}